\documentclass[final,3p,times]{elsarticle}
\usepackage{amssymb}
 \usepackage{amsthm}
 \usepackage{amsmath}
 \usepackage{graphicx}
\usepackage{color}
\usepackage{amsmath}%
\providecommand{\U}[1]{\protect\rule{.1in}{.1in}}

\begin{document}

\begin{frontmatter}



\title{Controlling the density of plasma species in $\rm {Ar/CF_4}$ Radiofrequency Capacitively Coupled Plasma Discharges}


\author{M.G. Elsheikh$^{1}$, Y. Abdella $^2$,T. El-Ashram $^1$,W.M. Moslem$^{1,2}$}
\author{M. Shihab$^{3}$}
\ead{mohammed.shihab@science.tanta.edu.eg}
\address{
$^1$ Department of Physics, Faculty of Science, Port Said University, 42521 Port Said,Egypt.\\
$^2$ Centre for Theoretical Physics, The British University in Egypt (BUE), El-Shorouk City, Cairo, Egypt.\\
$^3$ Tanta University, Faculty of Science, Physics Department,  Tanta 31527, Egypt.  }
%\fntext[label2]{$^*$mohammed.shihab@science.tanta.edu.eg}
%\author{}
%\address{Department of Physics, Faculty of Science, Benha University, Benha, P.O. Box 13518, Egypt.}
%\author{}
%\address{Ruhr-University Bochum, Applied Electrodynamics and Plasma Technology
%(AEPT), D-44780 Bochum, Germany}

\begin{abstract}

In this manuscript, a ﬂuid model is utilized to calculate the density of plasma species assuming geometrically symmetric $\rm Ar/CF_4$ Radiofrequency Capacitively Coupled Plasmas. The electrodes are driven by a sinusoidal wavefront with an amplitude of 200 V and a frequency of 13.56 MHz. The gap between the electrodes is 5 cm.The plasma species density is calculated as a function of the gas pressure, electron temperature, and gas composition. In good agreement with recent experimental results, $\rm {CF}^+_3$ and F are dominant for all considered simulation parameters. The results explain the pathways to perform atomic layer etching and nanolayer deposition processes.
In order to reveal the effect of electron heating on the discharge dynamics,  the spatio-temporal electron energy equation is coupled to the fluid model. Tailoring the driven potential has been found to control the concentration of some plasma species. When the plasma is driven with the fundamental frequency, Ohmic and stochastic heating allows electrons to be heated symmetrically. Higher harmonics give rise to an electrical asymmetry and electron heating asymmetry between the powered and grounded sheaths. The electron temperature depends on the driven harmonics; it adjusts gain and loss rates and some plasma species densities.
\end{abstract}

\begin{keyword}
${\rm Ar/CF_4}$ discharge, plasma species, dry etching, wet etching, deposition,  RF-CCPs.


\end{keyword}

\end{frontmatter}


%%%%%%%%%%%%%%%%%%%%%%
\section{Introduction}

The $\rm {Ar/CF_4}$ plasmas are indispensable for semiconductor and microelectronics  industries. They are a current topic for plasma simulation tools \cite{Dong}.
The atomic layer etching to produce nanostructures in semiconductor substrates is very important for micro and nanoelectronics industry. $\rm {CF_4}$ is one of the popular gases that is used in plasma etching \cite{Cardinaud,Lee,Horiike,Kanarik,Kikuchi}. The discharge of $CF_4$ is electronegative, negative ions are produced in the discharge, however, the electronegativity is weak at low pressures \cite{Aranka2015,Segawa1999}. Anisotropic etching may take place when positive ions are accelerated and hit substrates perpendicularly with sufficient energies. The etchant F atoms react with Si substrates yielding a volatile product in an isotropic etching process. In addition, a fluorinated silicon  – the product of the isotropic etching– has been found to deposit on  and forms 10 nm of a silicon layer in 4 min \cite{Akiki}. The discharge of $\rm{Ar/CF_4}$ is complex, tens of chemical reactions occur with different probabilities at the same time. Few experimental papers are devoted to measuring and estimating plasma species\cite{Li,Toneli}.

Capacitively coupled plasmas (CCPs) are simple and have been used frequently in material processing \cite{Chabert}. Different theoretical models have been devoted to understanding their discharge properties. Fluid models, Particle-in-Cell schemes as kinetic models, and the mix between fluid and kinetic models are used \cite{Mussenbrock,Zoheir}. The benchmark with experimental results shows that using these models properly may lead to the expected experimental results \cite{Kim,Shihab2021a,Prenzel2013}. Due to the large difference between the masses of electrons and heavy ions, usually, electrons gain energy first from the driven energy, then a part of the energy is transferred to the background gas by collision. At low pressure and for small collision rates, a non-equilibrium plasma is formed, with most of the medium -- ions and neutral background gas-- still at room temperature. This allows the treatment of substrates and biological samples without thermal effects. Moreover, free electrons are heated up to a few electronvolts. The heated free electrons undergo many chemical reactions and create different chemical species that could be used in material processing \cite{Lieberman} and plasma medicine \cite{Laroussi}. The rate of production of these chemical species depends on the electron temperature. Therefore, the electron heating mechanisms and modes are studied in different publications using different theoretical approaches, such as the confinement of beam electrons \cite{Wilczek}, $\alpha$-mode where the discharge is dominated by the collision of highly energetic  beam-like-electrons accelerated by the expanding sheaths with the plasma bulk, $\gamma$-mode where secondary electrons sustain excitation and ionization processes, a hybrid of both $\alpha$ and $\gamma$ modes at intermediate pressures \cite{Schulze2009}, and excitation of plasma series resonance \cite{Mussenbrock2006,Schungel2015a,Shihab2018,Elgendy2022}.

It is really important to reveal the effect of the discharge parameters on the species concentration. When a geometrically symmetric reactor is driven with a high enough radio-frequency voltage, symmetric discharge is formed. Two symmetric sheathes are generated close to the powered and grounded electrodes. Both sheaths have the same width, sheath potential, electron temperature, and the same rates of excitation and ionization. There is a phase shift between the two sheaths' dynamics, the collapse of one of them is the expansion of the other. The electron density and the electric field across the two sheaths will be displayed to show this phase shift. Also, the accumulated plasma species density in the whole discharge will be shown for various species. The density of $\rm {CF^+_3}$ and F species is greater than the density of other plasma species.
Electric asymmetry will rise when the discharge is driven by different radio frequencies. The powered sheath may be larger than the grounded sheath. Substrates on the powered electrode will be hit with energetic ions, and etching or sputtering may take place. On the grounded electrode where the sheath potential is small, a deposition process may take place. Therefore, tailoring the waveforms of the driven potential has been frequently used to optimize the ion energy distribution and the ion angular distribution in Radiofrequency Capacitively-Coupled Plasmas (RF-CCPs) via kinetic models \cite{Shihab2021a,Schungel2015,Shihab2017,Bruneau1,Bruneau2,Doyle,Sharma1,Sharma2}. Here, we would like to infer the consequences of tailoring the driven potential on the plasma species density and the electron temperature utilizing a fluid model. More chemical and atomic processes may be included in the fluid model and still the simulation computationally efficient.

The paper is organized as follows: The fluid model and its justification will be discussed in section (2). In section (3), the results of geometrically electrically symmetric results will be shown, then in section (4) the effect of using different harmonics on the electron temperature and the density of plasma species will be presented, and a justification using Particle-in-cell simulation. Finally, in section (5), a summary section will be given.

\section{The model}

In this contribution, we simulate $\rm {Ar/CF_4}$ discharges assuming a geometrically symmetric discharge with two planar electrodes separated by 5 cm. In the frame of a macroscopic description, the moments of the Boltzmann equation lead to conservation equations. The mass continuity equations where the rate of the plasma density is due to the unbalance between gain and loss rates is given as \cite{Shihab2021,Bai}:
\begin{equation}
\label{continuity_e_eq}
\frac{\partial{n_{\rm e}}}{\partial{t}}+\vec{\nabla} \cdot \vec{\Gamma}_{\rm e}=G_{\rm e}-L_{\rm e},
\end{equation}

\begin{equation}
\label{continuity_ip_eq}
\frac{\partial{n_{\rm ip}}}{\partial{t}}+\vec{\nabla} \cdot \vec{\Gamma}_{\rm ip}=G_{\rm ip}-L_{\rm ip},
\end{equation}

\begin{equation}
\label{continuity_in_eq}
\frac{\partial{n_{\rm in}}}{\partial{t}}+\vec{\nabla} \cdot \vec{\Gamma}_{\rm in}=G_{\rm in}-L_{\rm in},
\end{equation}

\begin{equation}
\label{continuity_m_eq}
\frac{\partial{n_{\rm m}}}{\partial{t}}+\vec{\nabla} \cdot \vec{\Gamma}_{\rm m}=G_{\rm m}-L_{\rm m},
\end{equation}
Here ${\rm e,ip,in,m}$ gives the species electrons, positive ions, negative ions, and neutral species, respectively. $n$ is the  density. ${\vec \Gamma}$ is the particles flux. $G$ and $L$ are gain and loss terms.

\begin{figure}
\center
\includegraphics[clip,width=0.6\linewidth]{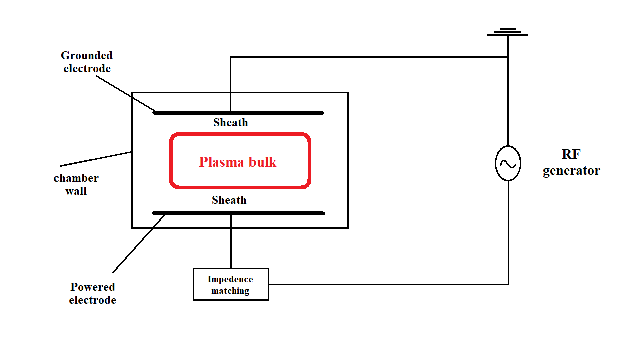}
\caption{A schematic for the plasma reactor}
\label{reactore}
\end{figure}

The first moment of the Boltzmann equation gives the conservation of momentum, where the rate of momentum is the summation of all forces. When the driving frequency is smaller than the collision frequency, the drift–diffusion (DD) approximation can be employed. In the DD approximation, the inertia of charged particles tends to zero, so that the mean velocities can be calculated as a function of the instantaneous electric field. The collision forces balance the electric forces. Also, the mean particle fluxes is the summation of a diffusion term due to spatial density gradients and a drift term for the charged particles due to the electric field. For a driven frequency of 13.56 MHz, the DD approximation for electrons holds for argon gas pressures greater than or equal to 100 mTorr \cite{Kim}. The particle fluxes are given as follows:

\begin{equation}
\label{momentum_e_eq}
\vec{\Gamma}_{\rm e} =- n_{\rm e} \,\, \mu_{\rm e} \,\, \vec{E}  -D_{\rm e}\vec{\nabla} n_{\rm e},
\end{equation}

\begin{equation}
\label{momentum_ip_eq}
\vec{\Gamma}_{\rm ip} =  n_{\rm ip} \,\, \mu_{\rm ip} \,\, \vec{E}  -D_{\rm ip}\vec{\nabla} n_{\rm ip},
\end{equation}

\begin{equation}
\label{momentum_in_eq}
\vec{\Gamma}_{\rm in} =  n_{\rm in} \,\, \mu_{\rm in} \,\, \vec{E}  -D_{\rm in}\vec{\nabla} n_{\rm in},
\end{equation}

and

\begin{equation}
\label{momentum_m_eq}
\vec{\Gamma}_{\rm m} = -D_{\rm m}\vec{\nabla} n_{\rm m}.
\end{equation}
 $\mu$ and $D$ are the species mobility and diffusion constants, respectively. $\vec{E}$ is the electric field.
 
 The second moment of the Boltzmann equation yields energy conservation. Due to the low ionization rate in low temperature plasmas and low pressure, the plasma is at room temperature, therefore,
\begin{equation}
 T_{\rm i}=T_{\rm m}=0.026 \,\, eV.
\end{equation}
$T$ is the temperature in energy units. The electron temperature could be assumed constant as
\begin{equation}\label{TeConstant}
 T_{\rm e}={\rm Constant} \,\, eV,
\end{equation}  
or the electron energy equation is coupled to the equation system as
\begin{equation}\label{TeEnergy}
 \frac{\partial n_{\rm e} T_{\rm e}}{\partial t}=-\vec{\nabla} \cdot (\frac{5}{3}  T_{\rm e} \vec{\Gamma}_{\rm e} -\frac{5}{3} n_{\rm e} D_{\rm e} \vec{\nabla} T_{\rm e})-e  \vec{\Gamma}_{\rm e} \cdot \vec{E}-n_{\rm e} n_{\rm G} k_{\rm loss},
\end{equation}
where $n_{\rm G}$ and $k_{\rm loss}$ are the gas density and the energy loss rate.
The three moments are closed with Poissons's equation to form a close set of equations:
\begin{equation}
 \nabla^2 \Phi=\frac{-e}{\epsilon_0} \left(\sum{n_{\rm ip}}-\sum{n_{\rm in}}-n_{\rm e}\right).
\end{equation}

 At the boundaries the particles flux is modified to include possible secondary emissions:
\begin{equation}
 \Gamma_{\rm ip}=\mu_{\rm ip} n_{\rm ip} E,
\end{equation}

\begin{equation}
 \Gamma_{\rm e}=k_{\rm s} n_{\rm e} - \gamma \Gamma_{\rm ip},
\end{equation}
where $k_s$ is the electron recombination coefficient and $\gamma$ is the secondary  electron emission coefficient. Secondary electron emission is not necessary for sustaining the discharge. Here we assumed $k_s=1.19\times 10^{7} {\rm cm/sec}$ and $\gamma=0.01$ \cite{Tagra}. The secondary emissions do not affect the results considerably.

There are different reaction rates in the literature. Here, we employ published reaction rates that are considered in \cite{Shihab2021} and in \cite{Bai} and references within it as \cite{Lee2005,Brok,Bogaerts,Bi,Mantzaris,Setareh,Schmidt,Zhang,Jackson}. The mobility and diffusion constants are assumed as given by \cite{So,Tagra,Kurihara}. Bolsig+ 2019 \cite{Hagelaar} is used to estimate rates of excitation and ionization and the energy loss coefficient ($k_{\rm loss}$) that are not given explicitly in \cite{Bai}; see the appendix.  Finally, as the area of the electrodes is larger than the gap size between the two electrodes and there is no external magnetic field, the model is simplified to be a 1D model. 

\section{Geometrically-electrically symmetric discharges}

\begin{figure}
\center
\includegraphics[clip,width=0.6\linewidth]{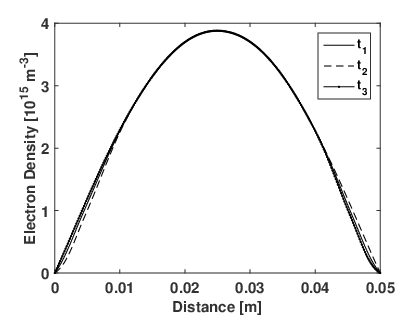}
\caption{The electron density at different phases inside the rf period ($T_{\rm rf}$): $t_1=T_{\rm rf}/4$, $t_2=T_{\rm rf}/2$, and $t_3=T_{\rm rf}$.}
\label{nedensity}
\end{figure}

\begin{figure}
\center
\includegraphics[clip,width=0.6\linewidth]{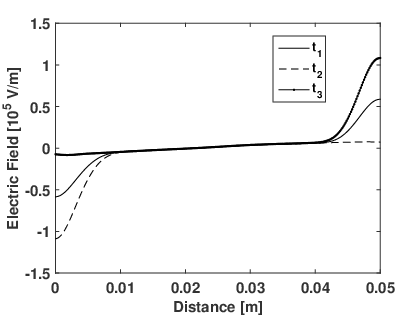}
\caption{The electric field at different phases inside the rf period ($T_{\rm rf}$): $t_1=T_{\rm rf}/4$, $t_2=T_{\rm rf}/2$, and $t_3=T_{\rm rf}$.}
\label{EField}
\end{figure}

 A schematic for the plasma reactor is displayed in \figurename{~\ref{reactore}}. The driven potential is 200 V, the driven frequency is 13.56 MHz, and the gap size is 5 cm. To avoid numerical instabilities, the time step is the RF periodic time ($T_{\rm rf}$) divided by 6000 and the gap size is divided into 299 intervals. The simulation runs until convergence, results do not change with the simulation time.
In \figurename{~\ref{nedensity}} and \figurename{~\ref{EField}}, we show the electron density and the electric field when the ratio of Ar is $50\%$, the ratio of $\rm {CF}_4$  is 50\%, and the electron temperature is constant and equals 2 eV. The model is solved employing a constant temperature using equation \ref{TeConstant} instead of \ref{TeEnergy}. The gas pressure is 200 mTorr. Electrons moves opposite to the electric field. Increasing the negative bias at the electrode allows sheath expansion and vice versa.


\begin{figure}
\center
\includegraphics[clip,width=0.6\linewidth]{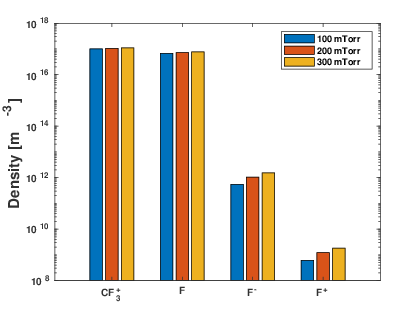}
\caption{The density of some plasma species for different pressures.}
\label{nsp1}
\end{figure}

In order to show the effect of changing the gas pressure, the gas composition, and the electron temperature, we repeated the simulation using the same other parameters. \figurename{~\ref{nsp1}}  shows the sum of densities between the two electrodes for some species as a function of the gas pressure. The F and $\rm{CF^+_3}$ dominate the discharge. Other species exist in the discharge, but their density is not more than  F and $\rm{CF^+_3}$. 

%\begin{figure}
%\center
%\includegraphics[clip,width=0.6\linewidth]{nsp2}
%\caption{The plasma species as a function of the discharge density for different %pressures.}
%\label{nsp2}
%\end{figure}

\begin{figure}
\center
\includegraphics[clip,width=0.6\linewidth]{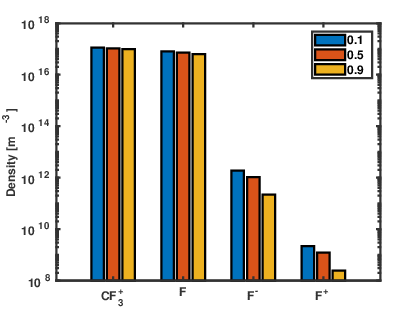}
\caption{The plasma species density as a function of Ar ratios.}
\label{nsr1}
\end{figure}

\begin{figure}
\center
\includegraphics[clip,width=0.6\linewidth]{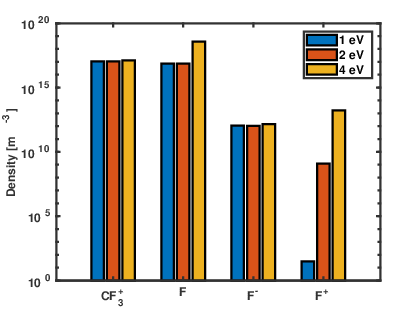}
\caption{The density of some plasma species at different temperatures.}
\label{nsT1}
\end{figure}

As shown in \figurename{~\ref{nsr1}}, by increasing the content of Ar to $\rm {CF_4}$, the density of $\rm {CF_4}$ species decreases, but still the density of F and $\rm{CF^+_3}$ larger than other species. 
Also, by changing the electron temperature, the species density changes as shown in \figurename{~\ref{nsT1}}. The density of some species is not sensitive to the electron temperature. This could be understood, not all reaction rates depend on electron-temperature.  The large densities of F and $\rm{CF^+_3}$ agree with measurements \cite{Li,Toneli}.

\section{Geometrically symmetric-electrically asymmetric discharges}

%\begin{figure}
%\center
%\includegraphics[clip,width=0.6\linewidth]{nsT2}
%\caption{The plasma species at different temperatures.}
%\label{nsp2}
%\end{figure}

\begin{figure}
\center
\includegraphics[clip,width=0.6\linewidth]{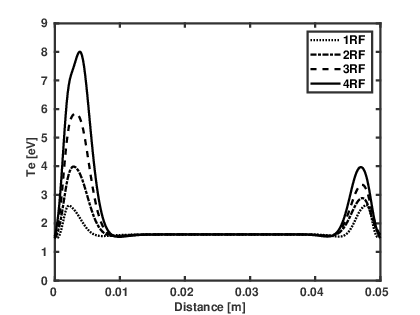}
\caption{The electron temperature when the plasma is generated via different harmonics. 1RF, 2RF, 3RF, and 4RF labeled the cases when N in equation \ref{harmonics} is 1,2,3, and 4, respectively.}
\label{Te}
\end{figure}

\begin{figure}
\center
\includegraphics[clip,width=0.6\linewidth]{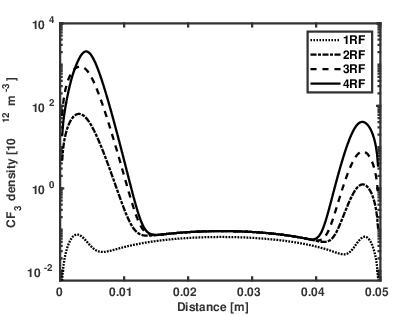}
\caption{The density of ${\rm CF_3}$ when the plasma is driven with different harmonics.}
\label{nCF3}
\end{figure}
\begin{figure}
\center
\includegraphics[clip,width=0.6\linewidth]{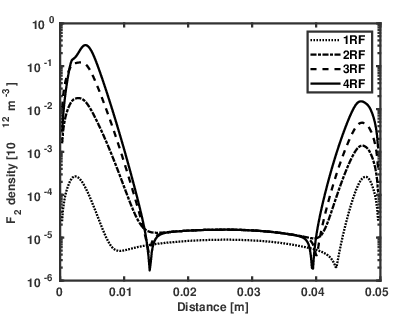}
\caption{The density of ${\rm F_2}$ when the plasma is driven with different harmonics.}
\label{nF2}
\end{figure}
\begin{figure}
\center
\includegraphics[clip,width=0.6\linewidth]{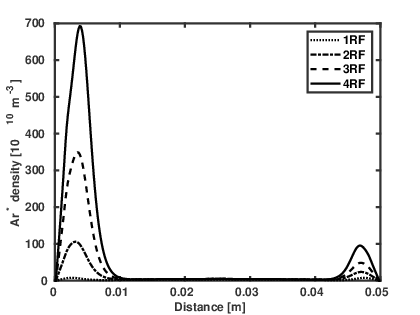}
\caption{The density of ${\rm Ar^*}$  when the plasma is driven with different harmonics.}
\label{Ar}
\end{figure}
\begin{figure}
\center
\includegraphics[clip,width=0.6\linewidth]{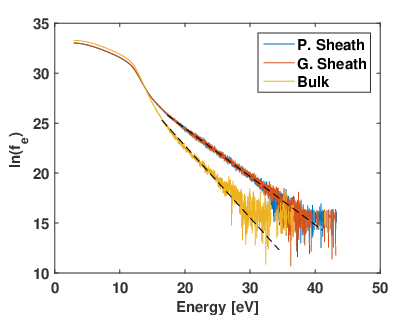}
\caption{The electron energy distribution in the powered sheath (P. Sheath) and grounded sheath (G. Sheath) and within the plasma bulk (Bulk). The discharge is driven with the fundamental frequency.}
\label{Te1RF}
\end{figure}
\begin{figure}
\center
\includegraphics[clip,width=0.6\linewidth]{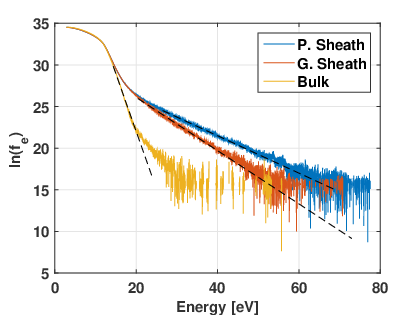}
\caption{The electron energy distribution in the powered sheath (P. Sheath) and grounded sheath (G. Sheath) and within the plasma bulk (Bulk).. The discharge is driven with three harmonics}
\label{Te3RF}
\end{figure}

Tailored voltage waveform is applied to capacitively coupled discharges to control the dc-self bias, the ion energy, and the ion flux \cite{Skarphedinsson,Schungel2016, Bruneau2016}. Tailored voltage waveform composed at least of  the fundamental frequency and the second harmonic. Generally, the driven voltage is a function of the frequency components, the amplitude of each frequency,  and the phase between them.
Here, the plasma is generated assuming consecutive harmonics, the driven potential is given as  
\begin{equation}\label{harmonics}
 V(t)=\sum_{k=1}^N V_0 \cos(2 \pi k f_0 t),
\end{equation}
where $V_0$ is the amplitude. The peak-to-peak amplitude is $2V_0=400$ Volt. $f_0$ is the fundamental frequency and equals $13.56$ MHz.
The simulation has been carried out employing different harmonics, in equation \ref{harmonics}, N=1,2,3, and 4. The electron energy equation is accounted for as given in  \ref{TeEnergy} instead of \ref{TeConstant}. The gap size is 5 cm, the Ar ratio is 0.9, and the gas pressure is 200 mTorr.
The temperature between the two electrodes is given in \figurename{~\ref{Te}}. 
The fundamental frequency produces symmetric discharge. Therefore, electrons are heated the same way in both sheaths \cite{Tagra,Tagra2}. 
%The density of different species as shown in Figures \ref{nCF3}-\ref{nF} is symmetric.
As seen in Figures \ref{nCF3}-\ref{Ar}, the density of different species produced by the fundamental frequency is symmetric. By adding more harmonics, an electrical asymmetry between the two sheaths arises. The electric field and the electrons flux within the powered sheath increase more than that within the grounded sheath. Consequently, electron heating and electron temperature are asymmetric. Ohmic and stochastic heating enhances  by increasing the applied potential due to the
increment of the sheath electric ﬁeld and electrons ﬂux.
Therefore, the driven harmonics could be used to adjust the density of plasma species as shown in  Figures \ref{nCF3}-\ref{Ar}.  
%Qualitatively similar behavior has been observed theoretically and experimentally for another gas mixture in RF-CCPs jet \cite{Korolov}. 
Particle-in-Cell (PIC) simulations under the same conditions considering only Ar species \cite{Shihab2017} show as in Figure \ref{Te1RF} that
the electron distributions within the powered sheath and the grounded sheath are the same when the discharge is driven with the fundamental frequency.
When three harmonics are considered, the electron energy distribution in the powered sheath differs than the electron energy distribution in the grounded sheath and within the plasma bulk; see in Figure \ref{Te3RF}. When imposing the Maxwell-Boltzmann distribution, the electron temperature is inversely proportional to the slope of the straight part. Dashed lines are added to \ref{Te1RF} and \ref{Te3RF} to guide the eye. As expected from our fluid calculations, in electric symmetric discharge, electrons in both sheaths are equal in temperature and hotter than the electrons in the plasma bulk. In electrically asymmetric discharge, electrons in the powered sheath  are hotter than that in the grounded sheath and within the plasma bulk. It is important to emphasize that one-to-one comparisons between fluid and kinetic results are difficult. Fluid dynamics produces macroscopic and averaged quantities. It is computationally efficient compared to kinetic calculations, but the current fluid model considers constant mobilities and diffusion coefficients. The findings of the fluid model could be improved by utlizing time dependent mobilities and diffusion coefficients extracted from PIC results.

\section{Summary}
The ${\rm Ar/CF_4}$ discharges are studied in this manuscript theoretically.  Two fluid model approaches have been used; one assumes a constant electron temperature in the entire discharge and the second considers the inhomogeneity of the electron temperature. The density of plasma species differs by changing the gas pressure, the gas composition, and the electron plasma temperature. Driven the discharge with different harmonics give rise to an electrical asymmetry, where in the powered sheath, electrons are heated more than in the ground sheath. Kinetic calculations via Particle-in-Cell confirms the findings of the fluid model. The electron temperature is a function of the driven harmonics. Therefore, the driven harmonics adjust the density of plasma species. As expected experimentally, our calculations predict the dominance of F and $\rm{CF^+_3}$ under typical discharge conditions.

\section{Acknowledgement}
M. Shihab acknowledges the Mission Department of the Ministry of Higher Education of Egypt for providing a scholarship to conduct this study and valuable discussions with Thomas Mussenbrock (Ruhr University Bochum). \\

\newpage
\vspace{100cm}
\section{Appendix}

\begin{table}[!htb]
\renewcommand{\arraystretch}{0.8}
    \caption{\it $\rm{Ar/CF_4}$ chemical reactions \cite{Bai,Lee2005,Brok,Bogaerts,Bi,Mantzaris,Setareh,Schmidt,Zhang,Jackson}. \label{table 1}}
    \centering
  \begin{tabular*}{\textwidth}{l @{\extracolsep{\fill}} lccc}
    \hline
Equation of Reaction& Rate of Reaction Coefficient $(m^3/s)$ \\
\hline

$e+Ar\rightarrow Ar^{+} + 2e$& $K_i=1.253\times 10^{-13}\exp({-18.618/T_e})$ $(m^3/s)$\\

$e+Ar\rightarrow Ar^* +e$& $K_{ex}=3.712\times 10^{-14}\exp({-15.06/T_e})$ $(m^3/s)$\\

$e+Ar^*\rightarrow Ar^{+}+2e$&$K_{si}=2.05\times 10^{-13}\exp({-4.95/T_e})$ $(m^3/s)$\\

$e+Ar^*\rightarrow Ar+e$&$K_{sc}=1.818\times 10^{-15}\exp({-2.14/T_e})$ $(m^3/s)$\\

$e+Ar^*\rightarrow Ar^r+e$&$K_{r}=2\times 10^{-13}$ $(m^3/s)$\\

$Ar^*+Ar^*\rightarrow Ar^{+}+Ar+e$&$K_{mp}=6.2\times 10^{-16}$ $(m^3/s)$\\

$Ar^*+Ar\rightarrow 2Ar$&$K_{2q}=3.0\times 10^{-21}$ $(m^3/s)$\\

$Ar^*+2Ar\rightarrow Ar+Ar_2$&$K_{3q}=1.1\times 10^{-42}$ $(m^6/s)$\\

$CF_3 + e\rightarrow CF_{3}^{+} + 2e$& $1.4\times 10^{-17}(11605 \times T_e)^{0.6481} \exp({-9.8/T_e})$\\

$F + e \rightarrow F^+ + 2e $ & $7.489 \times 10^{-19}(11605 \times T_e)^{0.8595}\exp({-17.6/T_e})$\\

$CF_4 + e \rightarrow CF_{3}^{+} + F + 2e$ & $1.159 \times 10^{-17}(11605 \times T_e)^{0.7645}\exp({-17.2/T_e})$\\

$CF_4 + e \rightarrow CF_{2}^{+} + F_2 + 2e$ & $2.886 \times 10^{-17}(11605 \times T_e)^{0.5108}\exp({-22.8/T_e})^*$ \\

$CF_4 + e \rightarrow CF^{+} + F_2 + F + 2e$ & $2.296 \times 10^{-20}(11605 \times T_e)^{1.09}\exp({-27/T_e})$\\

$CF_4 + e \rightarrow CF_3 + F^+ + 2e$ & $1.482 \times 10^{-19}(11605 \times T_e)^{0.9375}\exp({-34.7/T_e})$\\

$CF_4 + e \rightarrow CF_3  + F + e$ & $2 \times 10^{-15}\exp({-13/T_e})$\\

$CF_4 + e \rightarrow CF_2  + 2F + e$ & $5 \times 10^{-15}\exp({-13/T_e})$\\

$CF_3 + F^- \rightarrow CF_4  + e$ & $5 \times 10^{-16}$\\

$CF_2 + F^- \rightarrow CF_3  + e$ & $5 \times 10^{-16}$\\

$CF + F^- \rightarrow CF_2  + e$ & $5 \times 10^{-16}$\\

$CF_3 + F \rightarrow CF_4 $ & $2 \times 10^{-17}$\\

$CF_2 + F \rightarrow CF_3 $ & $1.3 \times 10^{-17}$\\

$CF + F \rightarrow CF2 $ & $5.2 \times 10^{-21}$\\

$CF_{3}^{-} + CF_{3}^{+} \rightarrow 2CF_3 $ & $4 \times 10^{-13}$\\

$F^- + CF_{3}^{+} \rightarrow F + CF_3$ & $4 \times 10^{-13}$\\

$F^- + CF_{2}^{+} \rightarrow F + CF_2$ & $1 \times 10^{-13} (11605 \times T_{i})^{-0.5}$\\

$F^- + CF^{+} \rightarrow F + CF$ & $1 \times 10^{-13} (11605 \times T_{i})^{-0.5}$\\

$F^- + F^{+} \rightarrow F_2$ & $4 \times 10^{-13} (11605 \times T_{i})^{-0.5}$\\

$CF_{2}^{+} + e \rightarrow CF +F$ & $4 \times 10^{-14}$\\

$CF^{+} + e \rightarrow C +F$ & $4 \times 10^{-14}$\\

$CF_{3}^{+} + e \rightarrow CF_3$ & $9.6 \times 10^{-13}$\\

$F^+ + e \rightarrow CF_3$ & $4 \times 10^{-14}$\\

$CF_4 + e \rightarrow CF_{3} + F^-$ & $4.8 \times 10^{-18}$\\

$CF_4 + e \rightarrow CF_{3}^{-} + F$ & $3.28 \times 10^{-17}$\\

$CF_2 + F_2 \rightarrow CF_3 + F$ & $4.56 \times 10^{-19}$\\

$CF_3 + F_2 \rightarrow CF_4 + F$ & $1.88 \times 10^{-20}$\\

$CF_{3}^{-} + F \rightarrow CF_3 + F^-$ & $ 5\times 10^{-14}$\\

$CF_{3}^{-} + Ar^+ \rightarrow CF_3 + Ar$ & $1 \times 10^{-13}$\\

$F^- + Ar^+ \rightarrow F + Ar$ & $1 \times 10^{-13}$\\

$CF_4 + Ar^+ \rightarrow CF_{3}^{+} + F + Ar$ & $9.58 \times 10^{-16}$\\

$Ar + CF_{3}^{+} \rightarrow CF_{3} + Ar^+$ & $1 \times 10^{-15}$\\
\hline
    \end{tabular*}
\end{table}
\end{document}